\documentclass[prd,twocolumn,showpacs,amsmath,amssymb,letterpaper]{revtex4}
\usepackage{graphicx}
\usepackage{bm}

\begin{document}

\title{Leptonic CP violation studies at MiniBooNE in the (3+2) sterile neutrino
 oscillation hypothesis}

\author{G.~Karagiorgi}
\email{georgia@nevis.columbia.edu}
\author{A.~Aguilar-Arevalo}
\email{alexis@phys.columbia.edu}
\author{J.~M.~Conrad}
\email{conrad@nevis.columbia.edu}
\author{M.~H.~Shaevitz}
\email{shaevitz@nevis.columbia.edu}
\affiliation{Department of Physics, Columbia University, New York, NY 10027}
\author{K.~Whisnant}
\email{whisnant@iastate.edu}
\affiliation{Department of Physics and Astronomy, Iowa State University, Ames, IA 50011}
\author{M.~Sorel}
\email{sorel@ific.uv.es}
\affiliation{Instituto de F\'{i}sica Corpuscular, IFIC, CSIC and Universidad de Valencia, Spain}
\author{V.~Barger}
\email{barger@physics.wisc.edu}
\affiliation{Department of Physics, University of Wisconsin, Madison, WI 53715}
\date{\today}

\begin{abstract}
We investigate the extent to which leptonic CP-violation in (3+2) sterile neutrino models leads
 to different oscillation probabilities for $\bar{\nu}_{\mu}\to\bar{\nu}_e$ and 
 $\nu_{\mu}\to\nu_e$ oscillations at MiniBooNE. We are using
 a combined analysis of short-baseline (SBL) oscillation
 results, including the LSND and null SBL results, to which we impose
 additional constraints from atmospheric oscillation data. We obtain
 the favored regions in MiniBooNE oscillation probability 
 space for both (3+2) CP-conserving and (3+2) CP-violating models.
 We further investigate the allowed CP-violation 
 phase values and the MiniBooNE reach for such a CP violation measurement.  
 The analysis shows that the oscillation probabilities in MiniBooNE neutrino
 and antineutrino running modes can differ significantly, with the latter
 possibly being as much as three times larger than the first. In addition, we also show
 that all possible values of the single CP-violation phase measurable at short
 baselines in (3+2) models are allowed within 99\% CL by existing data. 
\end{abstract}


\maketitle
\section{\label{sec:one}INTRODUCTION}
\begin{figure}[!tb]
\includegraphics*[width=4.0cm]{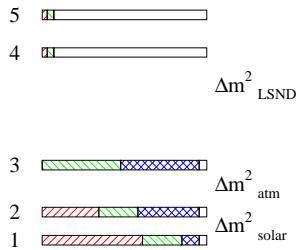}
 \caption{\label{epsfig:fig1}Flavor content of neutrino mass eigenstates
 in (3+2) models. Neutrino masses increase from
 bottom to top. The $\nu_e$ fractions are
 indicated by red (right-leaning) hatches, the $\nu_{\mu}$ fractions by
 green (left-leaning) hatches, the $\nu_{\tau}$ fractions by blue
 crosshatches, and the $\nu_s$ fractions by no hatches. The flavor contents
 shown are schematic only. \cite{Sorel:2003hf} }
\end{figure}

One of the most pressing open questions in neutrino physics today
 is whether or not leptons conserve the fundamental
 CP symmetry. The consequences of leptonic CP symmetry violation would
 be far-reaching and extend beyond the realm of particle physics,
 possibly being related to the matter-antimatter asymmetry observed in the
 Universe today \cite{Sakharov:1967dj}. \\
\indent In the standard paradigm of three-active-neutrino mixing occurring
 at the solar \cite{Cleveland:1998nv,sk_solar,Abdurashitov:2002nt,Hampel:1998xg,Altmann:2000ft,sno,Araki:2004mb} 
 and atmospheric \cite{sk_atmospheric,imb,macro,soudan-2,k2k,minos} oscillation 
 scales only, leptonic CP violation
 would yield different vacuum oscillation probabilities for
 neutrinos and antineutrinos that could be observed, for example, with
 accelerator-based neutrino oscillation appearance experiments operating near
 the atmospheric oscillation maximum
 \cite{Barger:2003qi,Mena:2005ek}. This is because CP-odd terms in the
 oscillation probability formula
 would appear from solar/atmospheric interference terms involving
 the single CP-violating Dirac phase appearing in the neutrino mixing 
 matrix \cite{Barger:1980jm}.\\
\indent Neutrino models involving active/sterile neutrino mixing
 \cite{Gomez-Cadenas:1995sj} at the
 LSND \cite{lsnd} neutrino mass splitting scale via at least two
 sterile neutrino states \cite{Peres:2000ic,Sorel:2003hf}
 would open the possibility for further manifestations of leptonic
 CP violation, including ones that could be measurable with neutrino
 appearance experiments at short baselines also. In this paper, we 
 investigate short-baseline (SBL) leptonic CP-violation in (3+2) sterile
 neutrino  models. A schematic diagram describing (3+2) sterile neutrino
 models is shown in Fig.~\ref{epsfig:fig1}. \\
\indent The analysis uses the same seven SBL datasets as in Ref.~\cite{Sorel:2003hf}, 
 including results on $\nu_{\mu}$ disappearance (from the CCFR84 \cite{Stockdale:1984cg}
 and CDHS \cite{Dydak:1983zq} experiments),
 $\nu_e$ disappearance (from the Bugey \cite{Declais:1994su} and CHOOZ
 \cite{Apollonio:2002gd} experiments), and
 $\nu_{\mu}\to\nu_e$ oscillations (from the LSND \cite{lsnd},
 KARMEN2 \cite{Armbruster:2002mp}, and NOMAD \cite{nomad}
 experiments). In addition, additional atmospheric 
 constraints have been added to the combined fit, based on the analysis of
 Ref.~\cite{Maltoni:2004gf}. \\
\indent Based on the combined analysis of the above SBL and atmospheric 
 oscillation data, we estimate the range of fundamental neutrino parameters 
 in (3+2) sterile neutrino models that are allowed within the 
 experimental capabilities of MiniBooNE, following a similar 
 analysis to that in Ref.~\cite{Sorel:2003hf}. However, in this case, the CP
 conservation requirement is relaxed, allowing for different neutrino and antineutrino oscillation
 probabilities.\\
\indent The paper is organized as follows. In Section \ref{sec:two},
 we specify the neutrino oscillation formalism used in this analysis to describe
 (3+2), CP-violating neutrino oscillations. In Section \ref{sec:three}, we
 discuss the analysis followed in this paper, used to constrain neutrino oscillation
 parameters with short-baseline and atmospheric data. In Section \ref{sec:four}, we present the results
 obtained for the CP-conserving, and CP-violating (3+2) models. 
 For both cases, we explore the oscillation probability asymmetry experimentally allowed in MiniBooNE, and
 we quote the neutrino mass and mixing parameters for the best-fit models derived
 from the combined SBL+atmospheric constraint analysis. In Section \ref{sec:five},
 we discuss the constraints on the single CP-violation phase 
 that could be measured at short baselines, inferred from a CP-violating (3+2)
 analysis of SBL+atmospheric oscillation results,
 and how the MiniBooNE CP asymmetry observable is expected to vary as a function 
 of this CP-violation phase.


\section{\label{sec:two}INCLUDING CP VIOLATION IN THE STERILE NEUTRINO OSCILLATION FORMALISM}
\indent In the sterile neutrino oscillation formalism, under the assumptions of CPT invariance, 
 the probability for a neutrino produced with flavor $\alpha$
 and energy $E$, to be detected as a neutrino of flavor $\beta$ after 
 traveling a distance $L$, is \cite{Barger:1999hi,Kayser:2002qs}:
\begin{eqnarray}
\label{eq:oscprob}
P(\nu_{\alpha}\to\nu_{\beta})= & \delta_{\alpha\beta}-4\sum_{i>j}\mathcal{R}(U^{\ast}_{\alpha i}
 U_{\beta i}U_{\alpha j}U^{\ast}_{\beta j})\sin^2x_{ij}+\nonumber \\
 & 2\sum_{i>j}\mathcal{I}(U^{\ast}_{\alpha i}U_{\beta i}U_{\alpha j}U^{\ast}_{\beta j})\sin2x_{ij}
\end{eqnarray}
\noindent where \begin{math}\mathcal{R}\end{math} and \begin{math}\mathcal{I}\end{math} 
 indicate the real and imaginary parts of the product of mixing matrix elements, 
 respectively; $\alpha ,\beta\equiv e,\mu ,\tau$, or $s$, ($s$ being the sterile flavor);
 $i,j=1,\ldots ,N$ ($N$ being the number of neutrino species); and 
 $x_{ij}\equiv 1.27\Delta m^2_{ij}L/E$. In defining $x_{ij}$, we take
 the neutrino mass splitting $\Delta m^2_{ij}\equiv m^2_i-m^2_j$
 in $\hbox{eV}^2$, the neutrino baseline $L$ in km, and the neutrino energy $E$ in GeV.
 For antineutrinos, the oscillation probability
 is obtained from Eq.~\ref{eq:oscprob} by replacing the mixing matrix
 $U$ with its complex-conjugate matrix. Therefore, if the mixing matrix is not real,
 neutrino and antineutrino oscillation probabilities can differ.\\
\indent For $N$ neutrino species, there are, in general, $(N-1)$ independent mass
 splittings, $N(N-1)/2$ independent moduli of parameters in the unitary
 mixing matrix, and $(N-1)(N-2)/2$ Dirac CP-violating phases that may be observed in
 oscillations. \\
\indent In SBL neutrino experiments that are sensitive
 only to $\nu_{\mu}\to\nu_{\not{\mu}}$, $\nu_e\to\nu_{\not{e}}$,
 and $\nu_{\mu}\to\nu_e$
 transitions, the set of observable parameters is reduced considerably.
 Firstly, oscillations due to atmospheric and solar mass splittings can
 be neglected in this case, or equivalently one can set
 $m_1=m_2=m_3$. Secondly, mixing matrix elements that measure the
 $\tau$ neutrino flavor fraction of the various neutrino mass
 eigenstates do not enter in the oscillation probability. In this case,
 the number of observable parameters is restricted to $(N-3)$ independent mass
 splittings, $2(N-3)$ moduli of mixing matrix parameters, and
 $N-4$ CP-violating phases. Therefore, for (3+2) sterile neutrino 
 models depicted in Fig.~\ref{epsfig:fig1}, that is for the $N=5$ case, 
 there are two independent mass splittings $\Delta m^2_{41}$ and $\Delta m^2_{51}$, 
 four moduli of mixing matrix parameters $|U_{e4}|,\ |U_{\mu 4}|,\ |U_{e5}|,\ |U_{\mu 5}|$,
 and one CP-violating phase. The convention used in the following for this CP-phase is:
\begin{equation}
\label{eq:cpvphase}
\phi_{45}=arg(U_{\mu 5}^*U_{e5}U_{\mu 4}U_{e4}^* )
\end{equation}
\noindent Under these assumptions, the general oscillation formula
 in Eq.~\ref{eq:oscprob} can be rewritten as:
\begin{eqnarray}
\label{eq:threeplustwo_1}
P(\nu_{\alpha}\to\nu_{\alpha}) = 1-4[(1-|U_{\alpha 4}|^2-|U_{\alpha 5}|^2)\cdot \nonumber\\
(|U_{\alpha 4}|^2\sin^2 x_{41}+|U_{\alpha 5}|^2\sin^2 x_{51})+ \nonumber \\
|U_{\alpha 4}|^2|U_{\alpha 5}|^2\sin^2 x_{54}] 
\end{eqnarray}
\noindent and
\begin{eqnarray}
\label{eq:threeplustwo_2}
P(\nu_{\mu}\to\nu_{e}) = 4|U_{\mu 4}|^2|U_{e 4}|^2\sin^2 x_{41}+ \nonumber \\
      4|U_{\mu 5}|^2|U_{e 5}|^2\sin^2 x_{51}+ \nonumber \\
      8|U_{\mu 5}||U_{e 5}||U_{\mu 4}||U_{e 4}|
 \sin x_{41}\sin x_{51}\cos (x_{54}-\phi_{45}) 
\end{eqnarray}
 The formulas for antineutrino oscillations are obtained by substituting
 $\phi_{45}\to -\phi_{45}$.


\section{\label{sec:three}ANALYSIS METHOD}
\indent The analysis we perform is a combined SBL+atmospheric analysis,
 with the purpose of obtaining the (3+2) model allowed regions
 in oscillation probability space for neutrino and anti-neutrino running 
 modes expected at MiniBooNE, and the allowed values of the CP-violation
 phase $\phi_{45}$. The physics and statistical assumptions used in the 
 analysis to describe the SBL experiments closely follow the  ones described 
 in detail in Ref.~\cite{Sorel:2003hf}.  The Monte Carlo method used 
 to apply the oscillation formalism discussed in Section \ref{sec:two}
 also closely follows the one described in Ref.~\cite{Sorel:2003hf}. The
 full set of oscillation parameters
 $(\Delta m^2_{41},|U_{e4}|,|U_{\mu 4}|,\Delta m^2_{51},|U_{e5}|,|U_{\mu 5}|,\phi_{45})$  
 is allowed to freely vary, constrained only by: 1) $0.1eV^2<\Delta m^2_{41},\Delta m^2_{51}<100eV^2$, with
 $\Delta m^2_{51}\ge\Delta m^2_{41}$, for definiteness; and 2) $|U_{ei}|^2+|U_{\mu i}|^2\le 0.5$, and 
 $|U_{\alpha 4}|^2+|U_{\alpha 5}|^2\le 0.5$. The first constraint defines
 the higher $\Delta m^2$ splitting range considered, imposed by the LSND
 signature, whereas the second constraint requires the fourth and fifth
 mass eigenstates to include only small active flavor quantities, as suggested
 by the solar and atmospheric oscillation data.\\
\indent A slight modification was made to the analysis method used in this
 paper, compared to the one used in Ref.~\cite{Sorel:2003hf}. Rather than
 generating neutrino masses and mixings in a random, unbiased way, we use
 importance sampling via a Markov chain Monte Carlo method \cite{braemaud,Metropolis:1953am},
 to better sample the regions in parameter space that provide a good fit to the
 SBL+atmospheric data. Given a starting point (model) $x_i$ in the 
 $(\Delta m^2_{41},|U_{e4}|,|U_{\mu 4}|,\Delta m^2_{51},|U_{e5}|,|U_{\mu 5}|,\phi_{45})$
 parameter  space, a trial state $x_{i+1}=x_i+e$ that depends only on the
 current state $x_i$ and on the probability distribution function 
 for the random vector $e$, is generated. The probability for the trial 
 state $x_{i+1}$ to be accepted as the new current state for further 
 model random generation is given by the transition probability: 
\begin{equation} 
\label{eq:transitionprobability} 
P(x_i\to x_{i+1})=min \{1,\exp [-(\chi^2_{i+1}-\chi^2_{i})/T]\} 
\end{equation} 
\noindent where $\chi^2_i$ and $\chi^2_{i+1}$ are $\chi^2$ values 
 for the states $x_i$ and $x_{i+1}$, quantifying the agreement 
 between the models and the short-baseline plus atmospheric 
 results used in the combined analysis, and $T$ is an effective 
 ``temperature'' parameter. The results presented here are obtained 
 by combining various Markov chains with different initial conditions, 
 probability distribution functions for $e$, and temperature parameters. 
 This modification allows for an efficient probe of the 
 larger dimensionality of the parameter set present 
 in CP-violating models, compared to CP-conserving models. \\
\indent The addition of atmospheric constraints to our previous
 analysis \cite{Sorel:2003hf} follows the assumptions discussed
 in Ref.~\cite{Maltoni:2004gf}. These constraints include 1489 days 
 of Super-Kamiokande charged-current data \cite{sk_atmospheric},
 including the $e$-like and $\mu$-like data samples of sub- and
 multi-GeV contained events, stopping events, and through-going
 upgoing muon data events. The analysis in Ref.~\cite{Maltoni:2004gf}
 assumes the three-dimensional atmospheric neutrino fluxes given in
 \cite{Honda:2004yz}, and a treatment of flux, cross-section, and
 experimental systematic uncertainties given in \cite{Gonzalez-Garcia:2004wg}.
 The atmospheric constraint includes also data on $\nu_{\mu}$ disappearance
 from the long baseline, accelerator-based experiment K2K \cite{k2k}. The
 atmospheric constraint is implemented in our analysis by simply adding a contribution
 $\chi^2_{\hbox{\tiny atm}}=\chi^2_{\hbox{\tiny atm}}(d_{\mu})$ to the total SBL contribution,
 $\chi^2_{\hbox{\tiny SBL}}=\chi^2_{\hbox{\tiny SBL}}(\Delta m^2_{41},|U_{e4}|,|U_{\mu 4}|,\Delta m^2_{51},|U_{e5}|,|U_{\mu 5}|,\phi_{45})$, 
 where:
\begin{equation}
\label{eq:dmu}
d_{\mu}=\frac{1-\sqrt{1-4A}}{2}
\end{equation}
\noindent with:
\begin{equation}
\label{eq:dmu2}
A\equiv (1-|U_{\mu 4}|^2-|U_{\mu 5}|^2)(|U_{\mu 4}|^2+|U_{\mu 5}|^2)+|U_{\mu 4}|^2|U_{\mu 5}|^2
\end{equation}
\noindent and by consequently adding a single degree of freedom to our analysis.
 We note that the recent analysis of atmospheric+K2K data in Ref.~\cite{Maltoni:2004gf}
 constrain the quantity $d_{\mu}$ in Eq.~\ref{eq:dmu}, and therefore muon neutrino
 disappearance at the LSND mass splitting scale, significantly more than previous
 results. For reference, the authors of Ref.~\cite{Maltoni:2004gf} quote an upper 
 limit on $d_{\mu}$ of 0.065 at 99\% C.L., while $d_{\mu}\le 0.13$ at 99\% C.L. is
 given in Ref.~\cite{Maltoni:2001mt}. \\
\indent The combined $\chi^2$ used to extract the best-fit values and allowed ranges
 for the fundamental oscillation parameters $\Delta m^2_{41}$, $\Delta m^2_{51}$,
 $U_{e4}$, $U_{\mu 4}$, $U_{e5}$, $U_{\mu 5}$, given in Sections \ref{sec:four}
 and \ref{sec:five}, is therefore:
\begin{equation} 
\label{eq:combinedchi2}
 \chi^2\equiv 
 \chi^2_{\hbox{\tiny SBL}}+\chi^2_{\hbox{\tiny atm}} 
\end{equation} 
\indent In CP-conserving models, $\phi_{45}$ is only allowed to take 
 values of $0$, or $\pi$, whereas in CP-violating models, $\phi_{45}$ can vary 
 within the full $(0,2\pi)$ range. Inclusion of this additional parameter
 reduces the total number of degrees of freedom by one.\\
\indent This analysis also provides realistic estimates of the oscillation
 probabilities to be expected in MiniBooNE in the framework of allowed
 CP-conserving and CP-violating (3+2) sterile neutrino 
 models. For that, expected neutrino transmutation rates for full
 $\nu_{\mu}\to\nu_e$ or $\bar{\nu}_{\mu}\to\bar{\nu}_e$ transmutations
 as a function of neutrino or anti-neutrino energy are considered, for
 neutrino and anti-neutrino running modes in MiniBooNE. These distributions
 are  weighted according to the oscillation probability formula in
 Eq.~\ref{eq:threeplustwo_2} to estimate the number of oscillation
 signal events for any (3+2) model, prior to event reconstruction and
 particle identification. The predictions for the full transmutation
 rates are obtained by multiplying the flux distributions as a function
 of energy for muon neutrinos and antineutrinos in both neutrino and antineutrino
 running modes (four flux distributions in total) by the (energy-dependent) total
 electron neutrino and antineutrino cross-sections on CH$_2$, respectively. The flux
 predictions are obtained from a full simulation of the FNAL Booster neutrino
 beamline \cite{MBrunplan}, while the neutrino cross-section predictions are
 obtained from the NUANCE event generator \cite{Casper:2002sd}. We do, therefore,
 take into account also the effect of ``wrong sign'' neutrinos in computing the
 expected oscillation probabilities, which have the effect of washing out 
 CP-violating observables. This effect is non-negligible since as much as one
 third of the total interaction rate in antineutrino running mode is expected
 to be due to neutrinos rather than antineutrinos; on the other hand, the
 antineutrino contribution in neutrino running mode is expected to be much smaller. 


\section{\label{sec:four}OSCILLATION PROBABILITY EXPECTATIONS FOR MINIBOONE}
\indent We define the oscillation probability in neutrino (anti-neutrino) mode
 expected at MiniBooNE as:
\begin{equation}
\label{eq:MBprob}
\stackrel{\hbox{\small{(-)}}}{~p}_{BooNE}=
\frac{
\int dE\ [p(\nu_{\mu}\to\nu_e)\stackrel{\hbox{\small{(-)}}}{N_0}(\nu )+
 p(\bar{\nu}_{\mu}\to\bar{\nu}_e)\stackrel{\hbox{\small{(-)}}}{N_0}(\bar{\nu})]
}
{
\int dE\ [\stackrel{\hbox{\small{(-)}}}{N_0}(\nu
)+\stackrel{\hbox{\small{(-)}}}{N_0}(\bar{\nu})]
}
\end{equation}
\noindent where $E$ is the neutrino energy; $p(\nu_{\mu}\to\nu_e)$ and $p(\nu_{\bar{\mu}}\to\nu_{\bar{e}})$
 are the oscillation probabilities given by Eq.~\ref{eq:threeplustwo_2}, with
 $\phi_{45}=0$ or $\pi$ for the CP-conserving case, and $0<\phi_{45}<2\pi$ 
 for the CP-violating case; $N_0(\nu )$ and $N_0(\bar{\nu})$ are the MiniBooNE neutrino and
 antineutrino full-transmutation rate distributions in neutrino
 running mode, and $\bar{N}_0(\nu )$ and $\bar{N}_0(\bar{\nu})$ are the neutrino and
 antineutrino full-transmutation rate distributions in antineutrino
 running mode, as defined in Section \ref{sec:three}. \\
\indent This section consists of two parts. In the first part, 
 we explore the experimentally allowed asymmetry in neutrino and antineutrino mode oscillation probabilities, $A_{p/\bar{p}}$,
 obtained from the SBL+atmospheric analysis assuming (3+2) CP-conserving models, as a function of the average oscillation
 probability allowed, $<p_{\hbox{\footnotesize BooNE}}>$. The asymmetry in oscillation probabilities and the average 
 oscillation probability are defined in Eq.~\ref{eq:cpasymmetry} and Eq.~\ref{eq:avgoscprob}, respectively.
\begin{equation}
\label{eq:cpasymmetry}
A_{p/\bar{p}} = \frac{p_{\hbox{\footnotesize BooNE}}-\bar{p}_{\hbox{\footnotesize BooNE}}}{p_{\hbox{\footnotesize BooNE}}+\bar{p}_{\hbox{\footnotesize BooNE}}}
\end{equation} 
\begin{equation}
\label{eq:avgoscprob}
<p_{\hbox{\footnotesize BooNE}}> = (p_{\hbox{\footnotesize BooNE}}+\bar{p}_{\hbox{\footnotesize BooNE}})/2
\end{equation}
 In the second part, we explore the allowed oscillation probability
 space $(p_{\hbox{\footnotesize BooNE}},\bar{p}_{\hbox{\footnotesize BooNE}})$ for (3+2) CP-violating
 models. In both parts, we quote the values for the masses and mixing parameters corresponding to
 the best-fit models in the $(p_{\hbox{\footnotesize BooNE}},\bar{p}_{\hbox{\footnotesize BooNE}})$ space.\\
\indent The 90\%, and 99\% CL allowed region are defined as the
 $(A_{p/\bar{p}},<p_{\hbox{\footnotesize BooNE}}>)$ or 
 $(p_{\hbox{\footnotesize BooNE}},\bar{p}_{\hbox{\footnotesize BooNE}})$ space for 
 which $\chi^2-\chi^2_{min}< 4.61$, and $\chi^2-\chi^2_{min}< 9.21$, respectively, 
 where $\chi^2_{min}$ is the absolute $\chi^2$ minimum for all
 $(p_{\hbox{\footnotesize BooNE}},\bar{p}_{\hbox{\footnotesize BooNE}})$ values. 
\subsection{\label{subsec:CPC}CP-conserving models results}
\begin{figure}[htb] 
 \includegraphics*[ width=\columnwidth, trim=10 20 0 0]{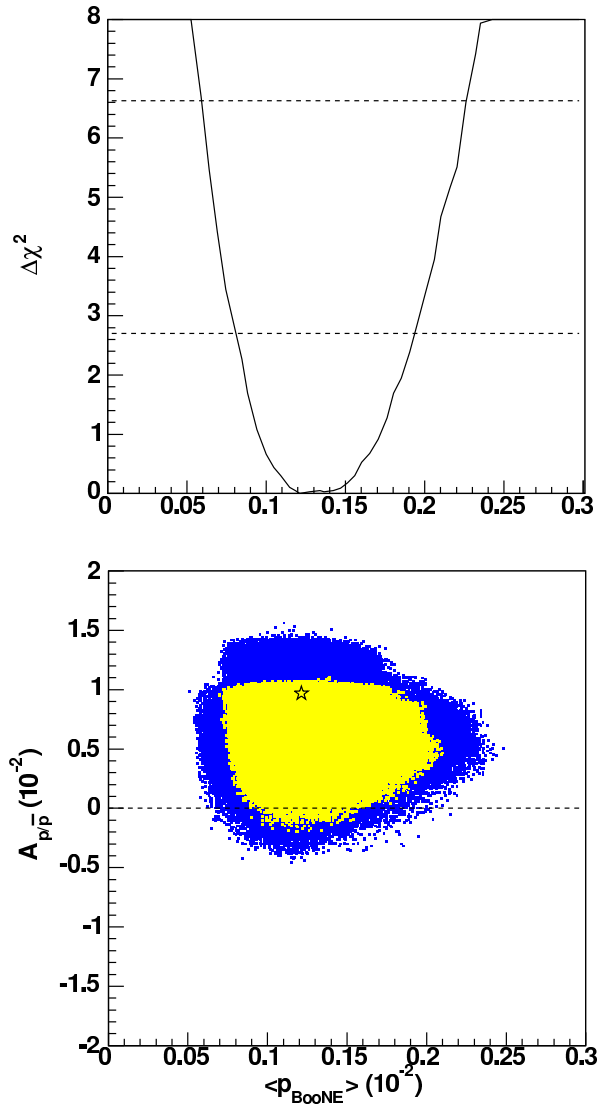} 
 \caption{\label{epsfig:fig2}Expected oscillation probability asymmetry at MiniBooNE for neutrino and 
 antineutrino running modes, for CP-conserving (3+2) models. The yellow (light gray) region corresponds to the 90\% CL allowed region;
 the blue (dark gray) region corresponds to the 99\% CL allowed region. See text for details.} 
\end{figure}

\indent Fig.~\ref{epsfig:fig2} shows predictions
 for the asymmetry in oscillation probabilities expected in MiniBooNE
 neutrino and antineutrino modes, in the CP-conserving,
 (3+2) sterile neutrino hypothesis. 
\indent The bottom panel in Fig.~\ref{epsfig:fig2}
 shows the region
 in $(A_{p/\bar{p}},<p_{\hbox{\footnotesize BooNE}}>)$ space that is allowed at the
 90\% and 99\% confidence level (2 dof) by existing short-baseline data used in 
 the analysis, including LSND. The star indicates the best-fit, at
 $p_{\hbox{\footnotesize BooNE}}\simeq \bar{p}_{\hbox{\footnotesize BooNE}}\simeq 0.13\cdot 10^{-2}$.
 The effect of ``fake'' CP-violation due to spectrum  differences in neutrino and
 antineutrino running modes manifests itself as a departure from zero-asymmetry indicated by the dotted line in the bottom panel
 of Fig.~\ref{epsfig:fig2}. The effect is at the percent level
 at most. \\
\indent The top panel in
 Fig.~\ref{epsfig:fig2} shows the 1-dimensional projection
 of the $\Delta\chi^2$ contour obtained from the SBL+atmospheric data,
 as a function of average oscillation probability $<p_{\hbox{\footnotesize BooNE}}>$.
 The dashed lines at $\Delta \chi^2=2.70$ and $6.63$ indicate the 90\% and 99\%
 CL regions, respectively (1 dof). MiniBooNE is expected to
 measure an oscillation probability in excess of $\simeq 0.05\cdot
 10^{-2}$ if CP-conserving (3+2) models are correct. \\
\indent The best-fit model parameters for CP-conserving (3+2) sterile 
 neutrino oscillation models are shown in Table \ref{tab:bestfitmodelscpcvscpv}.
\subsection{\label{subsec:CPV}CP-violating models results}
\begin{figure}[htb]
 \includegraphics*[ width=\columnwidth, trim=0 45 20 0, angle=-90]{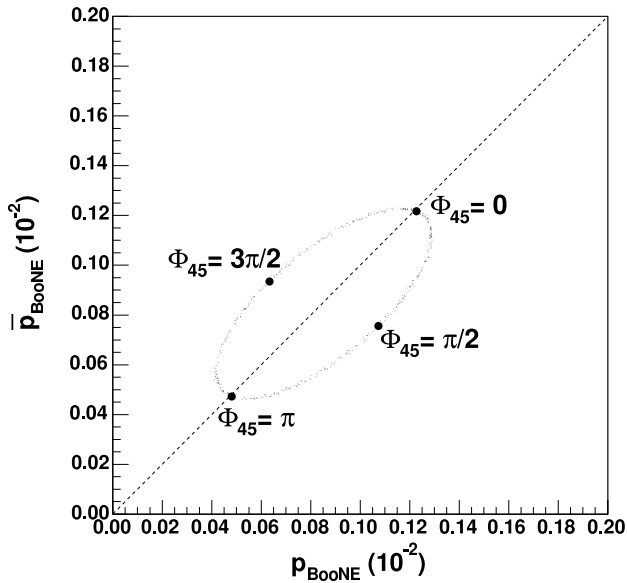}
 \caption{\label{epsfig:fig3} Illustration of expected oscillation probabilities at MiniBooNE in neutrino and 
 antineutrino running modes, for CP-violating (3+2) models with atmospheric
 constraint. Here, the neutrino masses and mixings are fixed to their best-fit values  
 and the only parameter that is allowed to vary is the CP-violating phase, $\phi_{45}$.}
\end{figure}

\begin{figure}[htb]
 \includegraphics*[width=\columnwidth, trim=30 30 0 0, angle=-90]{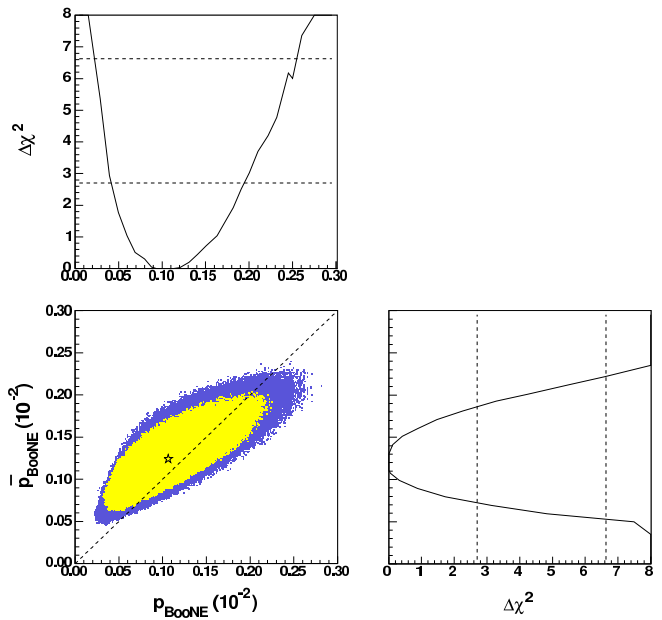}
 \caption{\label{epsfig:fig4}Expected oscillation probabilities at MiniBooNE in neutrino and
 antineutrino running modes, for CP-violating (3+2) models. The yellow (light gray) region corresponds to the 90\% CL allowed region; 
 the blue (dark gray) region corresponds to the 99\% CL allowed region. See text for details.}
\end{figure}

\indent Fig.~\ref{epsfig:fig3} shows the order of magnitude of the
 CP-violating effects to be expected in
 $(p_{\hbox{\footnotesize BooNE}},\bar{p}_{\hbox{\footnotesize BooNE}})$ space, 
 as $\phi_{45}$ is varied over its allowed range $(0,2\pi)$, while the
 remaining oscillation parameters are fixed to their best-fit values for the CP-conserving case. In this
 particular Figure, where we neglect goodness-of-fit considerations of the
 SBL+atmospheric datasets as a function of $\phi_{45}$ variations,
 neutrino/anti-neutrino oscillation probability differences as large as a factor of two can be obtained, near maximal
 CP-violation ($\phi_{45}=\pi/2$ or $3\pi/2$). As will be seen below, a similar conclusion (with actually
 even larger differences allowed among neutrino and antineutrino running
 modes) is reached with a more quantitative analysis that takes into account
 $\chi^2$ variations as a function of all neutrino parameters. \\
\indent Fig.~\ref{epsfig:fig4} shows the oscillation probabilities to be expected at MiniBooNE in
 neutrino and anti-neutrino running modes, in a CP-violating, (3+2) scenario. \\ 
\noindent Unlike in Fig.~\ref{epsfig:fig3}, in Fig.~\ref{epsfig:fig4}
 all parameters ($\Delta m^2_{41}$, $\Delta m^2_{51}$, $|U_{e4}|$, $|U_{\mu 4}|$, $|U_{e5}|$, $|U_{\mu 5}|$, $\phi_{45}$) are now allowed to 
 vary within the constraints provided by existing SBL+atmospheric oscillation results. Compared to the CP-conserving case of
 Fig.~\ref{epsfig:fig2}, the best-fit point (indicated by a star) does not change significantly; however, large
 asymmetries in oscillation probability due to CP-violation are now possible,
 shown by departures from the dashed line in the bottom left panel of Fig.~
 \ref{epsfig:fig4}. The general trend is that the 2-dimensional allowed region in 
 $(p_{\hbox{\footnotesize BooNE}},\bar{p}_{\hbox{\footnotesize BooNE}})$ 
 space is tilted more horizontally compared to the dashed 
 line $\bar{p}_{\hbox{\footnotesize BooNE}}=p_{\hbox{\footnotesize BooNE}}$, indicating that existing 
 short-baseline results constrain more $\bar{\nu}_{\mu}\to\bar{\nu}_e$ 
 than $\nu_{\mu}\to\nu_e$ oscillations. \\ 
\indent The best-fit model parameters for CP-violating (3+2) sterile  
 neutrino oscillation models are shown in Table
 \ref{tab:bestfitmodelscpcvscpv}. From a comparison of the $\chi^2$ values
 given in the Table, it is clear that CP-violating, (3+2) models do not
 provide a significantly better description of short-baseline and atmospheric
 data, compared to CP-conserving, (3+2) models.
\begingroup
\squeezetable
\begin{table}[bt] 
\begin{ruledtabular} 
\begin{tabular}{ccccllllr} 
 Model & $\chi^2\ (d.o.f.)$ & $\Delta m^2_{41}$ & $\Delta m^2_{51}$ & $|U_{e4}|$ & $|U_{\mu4}|$ & $|U_{e5}|$ & $|U_{\mu5}|$ & $\phi_{45}$ \\ \hline 
 CPC & 141.4 (145) & 0.92 & 24 & 0.132 & 0.158 & 0.066 & 0.159 & 0 \\
 CPV & 140.8 (144) & 0.91 & 24 & 0.127 & 0.147 & 0.068 & 0.164 & 1.8$\pi$\\
\end{tabular} 
\end{ruledtabular} 
\caption{\label{tab:bestfitmodelscpcvscpv} Comparison of best-fit values for mass-splittings and mixing parameters for   
 (3+2) CP-conserving and CP-violating models. Mass splittings are shown in $eV^2$. See text for details.} 
\end{table} 
\endgroup


\section{\label{sec:five}CONSTRAINTS ON CP-VIOLATION PHASE}
\indent In this section we discuss the  present constraints on the short-baseline, CP-violating phase $\phi_{45}$ that the
 current SBL+atmospheric oscillation data impose on (3+2) sterile neutrino oscillation models,
 and the prospects of observing such phase at MiniBooNE.
\begin{figure}[!tb] 
 \includegraphics*[width=\columnwidth, trim=30 30 0 0, angle=-90]{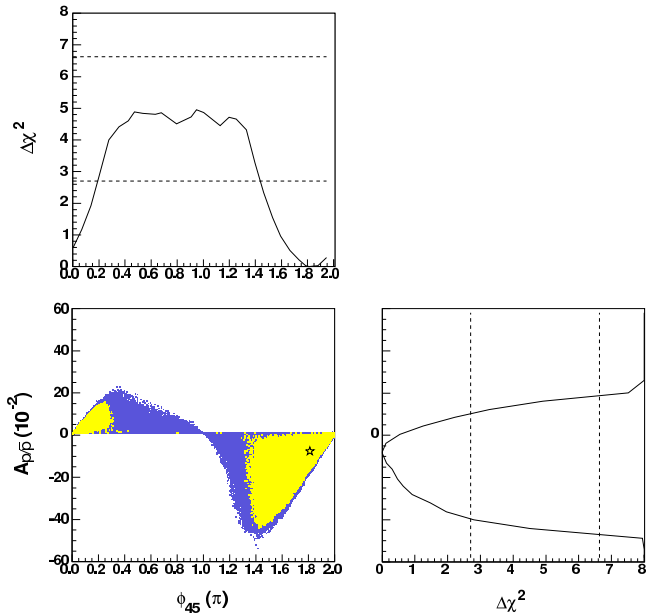}
 \caption{\label{epsfig:fig5}Current limits on the CP-violating phase $\phi_{45}$ 
 from current short-baseline results, and CP asymmetry measurement expected 
 at MiniBooNE, $A_{p/\bar{p}}$, as a function of $\phi_{45}$. The yellow (light gray) region corresponds to the 90\% CL allowed region;
 the blue (dark gray) region corresponds to the 99\% CL allowed region. See text for details.} 
\end{figure} 
\indent The top left panel in Fig.~\ref{epsfig:fig5} 
 shows that all values for the CP-phase $\phi_{45}$ are presently 
 allowed at the 99\% confidence level, and that CP-violating, 
 (3+2) models with small degrees of CP violation are marginally 
 preferred.  The bottom left plot shows that large CP asymmetry is 
 possible, but not required, for maximal CP-violation, given by phases 
 of around $\phi_{45}=\pi/2$ and $3\pi /2$. In particular, CP asymmetries
 up to $A_{p/\bar{p}}\simeq -0.5$ could be obtained,
 where $A_{p/\bar{p}}$ is defined in 
 Eq.~\ref{eq:cpasymmetry}. The value $A_{p/\bar{p}}=-0.5$ corresponds to three times larger oscillation 
 probability in antineutrino running mode, compared to the neutrino running mode  
 probability. Comparing Fig.~\ref{epsfig:fig5} with Fig.~\ref{epsfig:fig2}, we conclude that a significant 
 departure from zero in the asymmetry observable $A_{p/\bar{p}}$ could naturally 
 be interpreted as a manifestation of leptonic CP violation.


\section{\label{sec:six}CONCLUSIONS}
\indent We have performed a combined analysis of data from seven
 short-baseline experiments (Bugey, CHOOZ, CCFR84, CDHS, KARMEN,
 LSND, NOMAD), and including also constraints from atmospheric oscillation
 data (Super-Kamiokande, K2K), for the (3+2) neutrino oscillation hypothesis,
 with two sterile neutrinos at high $\Delta m^2$. The motivation for 
 considering two light, sterile neutrino models arises from the tension in
 trying to reconcile, in a CPT-conserving framework, the LSND
 signal for oscillations with the null results obtained by the other SBL
 experiments with a single light sterile neutrino state
 \cite{Sorel:2003hf,Maltoni:2002xd,Strumia:2002fw,Grimus:2001mn}.
 The class of (3+2) sterile neutrino models open up the possibility
 of observing possible leptonic CP-violating effects at short-baseline experiments, and in
 particular within the experimental capabilities of MiniBooNE.\\
\indent We have described two types of analyses in the (3+2) neutrino oscillation hypothesis.
 In the first analysis, we treat the SBL datasets with additional atmospheric constraints
 in a CP-conserving scenario, and we determine the allowed oscillation probabilities
 at MiniBooNE in both neutrino and antineutrino running modes, as well as the best-fit
 values for the mass splittings and mixing parameters. In the second analysis, we
 consider a CP-violating scenario to obtain the favored regions in MiniBooNE oscillation
 probability space, we determine the best-fit values for the mass splittings
 and mixing parameters, and we further investigate the allowed 
 CP-violating phase values, quoting the best-fit value for the CP-violating phase. \\
\indent The main results of the analysis are given in Sections
 \ref{sec:four} and \ref{sec:five}. First, we find that
 CP-violating, (3+2) models do not provide a significantly better
 description of short-baseline and atmospheric data, compared to
 CP-conserving, (3+2) models. On the other hand, even if only a small degree
 of CP violation is marginally preferred, we also find that existing
 data allow for all possible values for the single CP-violating phase that
 could be observed at short baselines in (3+2) models, at 99\% C.L..
 Finally, if leptonic CP violation occurs and (3+2) sterile neutrino models
 are a good description of the data, we find that differences as large as a
 factor of three between the electron (anti-)neutrino appearance probabilities
 in neutrino and antineutrino running modes at MiniBooNE are possible.\\
\indent The existence of a fifth neutrino with mass of order 5~eV, as found in 
 our fits, would be in conflict with cosmological bounds obtained under 
 the assumption that all the neutrinos are in thermal 
 equilibrium, see, e.\-g.\-, \cite{Hannestad:2005ey}. However, these bounds may be 
 avoided if the neutrinos do not thermalize \cite{Abazajian:2004aj} or if 
 the reheating temperature of the universe is very low \cite{Gelmini:2004ah}.


\begin{acknowledgments}
\noindent We thank O.~Yasuda for useful suggestions, M.~Maltoni for providing the
 data on atmospheric constraints, the MiniBooNE
 Collaboration for providing the neutrino flux and cross-section expectations
 for the MiniBooNE experiment, K.~Abazajian for introducing one of us (MS)
 to Markov chain Monte Carlo simulations, and S.~Parke for enlightening discussions and for 
 pointing out a mistake in an earlier version of this paper. This work was supported by NSF grant no.~PHY-0500492,
 by U.S. Dept. of Energy grant no.\-s DE-FG02-95ER40896 and DE-FG02-01ER41155, by the Wisconsin
 Alumni Research Foundation, and by a Marie Curie Intra-European
 Fellowship within the 6th European Framework Program.  
\end{acknowledgments}


\newpage


\begin{thebibliography}{99}
\bibitem{Sakharov:1967dj}
  A.~D.~Sakharov,
  Pisma Zh.\ Eksp.\ Teor.\ Fiz.\  {\bf 5}, 32 (1967)
  [JETP Lett.\  {\bf 5}, 24 (1967\ SOPUA,34,392-393.1991\ UFNAA,161,61-64.1991)].
\bibitem{Cleveland:1998nv}
  B.~T.~Cleveland {\it et al.},
  Astrophys.\ J.\  {\bf 496}, 505 (1998).
\bibitem{sk_solar}
  Y.~Fukuda {\it et ll.}  [Super-Kamiokande Collaboration],
  Phys.\ Rev.\ Lett.\  {\bf 81}, 1158 (1998)
  [Erratum-ibid.\  {\bf 81}, 4279 (1998)]
  [arXiv:hep-ex/9805021];
%
  Phys.\ Rev.\ Lett.\  {\bf 82}, 2430 (1999)
  [arXiv:hep-ex/9812011].
\bibitem{Abdurashitov:2002nt}
  J.~N.~Abdurashitov {\it et al.}  [SAGE Collaboration],
  J.\ Exp.\ Theor.\ Phys.\  {\bf 95}, 181 (2002)
  [Zh.\ Eksp.\ Teor.\ Fiz.\  {\bf 122}, 211 (2002)]
  [arXiv:astro-ph/0204245].
\bibitem{Hampel:1998xg}
  W.~Hampel {\it et al.}  [GALLEX Collaboration],
  Phys.\ Lett.\ B {\bf 447}, 127 (1999).
\bibitem{Altmann:2000ft}
  M.~Altmann {\it et al.}  [GNO Collaboration],
  Phys.\ Lett.\ B {\bf 490}, 16 (2000)
  [arXiv:hep-ex/0006034].
\bibitem{sno}
  Q.~R.~Ahmad {\it et al.}  [SNO Collaboration],
  Phys.\ Rev.\ Lett.\  {\bf 89}, 011301 (2002)
  [arXiv:nucl-ex/0204008];
%
  Phys.\ Rev.\ Lett.\  {\bf 89}, 011302 (2002)
  [arXiv:nucl-ex/0204009];
%
  Phys.\ Rev.\ Lett.\  {\bf 87}, 071301 (2001)
  [arXiv:nucl-ex/0106015].
\bibitem{Araki:2004mb}
  T.~Araki {\it et al.}  [KamLAND Collaboration],
  Phys.\ Rev.\ Lett.\  {\bf 94}, 081801 (2005)
  [arXiv:hep-ex/0406035].
\bibitem{sk_atmospheric}
  Y.~Fukuda {\it et al.}  [Super-Kamiokande Collaboration],
  Phys.\ Lett.\ B {\bf 433}, 9 (1998)
  [arXiv:hep-ex/9803006];
%
  Phys.\ Lett.\ B {\bf 436}, 33 (1998)
  [arXiv:hep-ex/9805006];
%
  Phys.\ Rev.\ Lett.\  {\bf 81}, 1562 (1998)
  [arXiv:hep-ex/9807003];
%
  Phys.\ Rev.\ Lett.\  {\bf 82}, 2644 (1999)
  [arXiv:hep-ex/9812014];
%
  Phys.\ Lett.\ B {\bf 467}, 185 (1999)
  [arXiv:hep-ex/9908049];
%
  Y.~Ashie {\it et al.}  [Super-Kamiokande Collaboration],
  Phys.\ Rev.\ D {\bf 71}, 112005 (2005)
  [arXiv:hep-ex/0501064].
\bibitem{imb}
  D.~Casper {\it et al.},
  Phys.\ Rev.\ Lett.\  {\bf 66}, 2561 (1991);
%
  R.~Becker-Szendy {\it et al.},
  Phys.\ Rev.\ Lett.\  {\bf 69}, 1010 (1992).
\bibitem{macro}
  M.~Ambrosio {\it et al.}  [MACRO Collaboration],
  Phys.\ Lett.\ B {\bf 434}, 451 (1998)
  [arXiv:hep-ex/9807005];
%
  Phys.\ Lett.\ B {\bf 478}, 5 (2000)
  [arXiv:hep-ex/0001044];
%
  Phys.\ Lett.\ B {\bf 517}, 59 (2001)
  [arXiv:hep-ex/0106049];
%
  Phys.\ Lett.\ B {\bf 566}, 35 (2003)
  [arXiv:hep-ex/0304037].
\bibitem{soudan-2}
  W.~W.~M.~Allison {\it et al.},
  Phys.\ Lett.\ B {\bf 391}, 491 (1997)
  [arXiv:hep-ex/9611007];
%
  Phys.\ Lett.\ B {\bf 449}, 137 (1999)
  [arXiv:hep-ex/9901024];
%
  M.~C.~Sanchez {\it et al.}  [Soudan 2 Collaboration],
  [arXiv:hep-ex/0307069].
\bibitem{k2k}
  S.~H.~Ahn {\it et al.}  [K2K Collaboration],
  Phys.\ Lett.\ B {\bf 511}, 178 (2001)
  [arXiv:hep-ex/0103001];
%
  Phys.\ Rev.\ Lett.\  {\bf 90}, 041801 (2003)
  [arXiv:hep-ex/0212007];
%
  Phys.\ Rev.\ D {\bf 74}, 072003 (2006) 
  [arXiv:hep-ex/0606032].
\bibitem{minos}
  D.~G.~Michael {\it et al.} [MINOS Collaboration],
  [arxiv:hep-ex/0607088].
\bibitem{Barger:2003qi}
  V.~Barger, D.~Marfatia and K.~Whisnant,
  Int.\ J.\ Mod.\ Phys.\ E {\bf 12}, 569 (2003)
  [arXiv:hep-ph/0308123].
\bibitem{Mena:2005ek}
  O.~Mena,
  Mod.\ Phys.\ Lett.\ A {\bf 20}, 1 (2005)
  [arXiv:hep-ph/0503097].
\bibitem{Barger:1980jm}
  V.~Barger, K.~Whisnant and R.~J.~N.~Phillips,
  Phys.\ Rev.\ Lett.\ {\bf 45}, 2084 (1980).
\bibitem{Gomez-Cadenas:1995sj}
  J.~J.~Gomez-Cadenas and M.~C.~Gonzalez-Garcia,
  Z.\ Phys.\ C {\bf 71}, 443 (1996)
  [arXiv:hep-ph/9504246].
\bibitem{lsnd}
  C.~Athanassopoulos {\it et al.}  [LSND Collaboration],
  Phys.\ Rev.\ Lett.\  {\bf 77}, 3082 (1996)
  [arXiv:nucl-ex/9605003];
%
  C.~Athanassopoulos {\it et al.}  [LSND Collaboration],
  Phys.\ Rev.\ C {\bf 58}, 2489 (1998)
  [arXiv:nucl-ex/9706006];
%
  A.~Aguilar {\it et al.}  [LSND Collaboration],
  Phys.\ Rev.\ D {\bf 64}, 112007 (2001)
  [arXiv:hep-ex/0104049].
\bibitem{Peres:2000ic}
  O.~L.~G.~Peres and A.~Y.~Smirnov,
  Nucl.\ Phys.\ B {\bf 599}, 3 (2001)
  [arXiv:hep-ph/0011054].
\bibitem{Sorel:2003hf} 
  M.~Sorel, J.~M.~Conrad and M.~H.~Shaevitz, 
  Phys.\ Rev.\ D {\bf 70}, 073004 (2004)  
  [arXiv:hep-ph/0305255]. 
\bibitem{Stockdale:1984cg}
  I.~E.~Stockdale {\it et al.},
  Phys.\ Rev.\ Lett.\  {\bf 52}, 1384 (1984).
\bibitem{Dydak:1983zq}
  F.~Dydak {\it et al.},
  Phys.\ Lett.\ B {\bf 134}, 281 (1984).
\bibitem{Declais:1994su}
  Y.~Declais {\it et al.},
  Nucl.\ Phys.\ B {\bf 434}, 503 (1995).
\bibitem{Apollonio:2002gd}
  M.~Apollonio {\it et al.},
  [arXiv:hep-ex/0301017].
\bibitem{Armbruster:2002mp}
  B.~Armbruster {\it et al.}  [KARMEN Collaboration],
  Phys.\ Rev.\ D {\bf 65}, 112001 (2002)
  [arXiv:hep-ex/0203021].
\bibitem{nomad}
  P.~Astier {\it et al.}  [NOMAD Collaboration],
  Phys.\ Lett.\ B {\bf 570}, 19 (2003)
  [arXiv:hep-ex/0306037];
%
  D.~Gibin,
  Nucl.\ Phys.\ Proc.\ Suppl.\  {\bf 66}, 366 (1998);
%
  V.~Valuev  [NOMAD Collaboration],
  \href{http://www.slac.stanford.edu/spires/find/hep/www?irn=4920686}{SPIRES entry}
  {\it Prepared for International Europhysics Conference on High-Energy Physics (HEP 2001), Budapest, Hungary, 12-18 Jul 2001}
\bibitem{Maltoni:2004gf}
  M.~Maltoni, T.~Schwetz, M.~Tortola and J.~W.~F.~Valle,
  New J.\ Phys.\ {\bf 6}, 122 (2004)
  [arXiv:hep-ph/0405172].
\bibitem{Barger:1999hi}
  V.~Barger, Y.~B.~Dai, K.~Whisnant and B.~L.~Young,
  Phys.\ Rev.\ D {\bf 59}, 113010 (1999)
  [arxiv:hep-ph/9901388].
\bibitem{Kayser:2002qs} 
  B.~Kayser, 
  [arXiv:hep-ph/0211134]. 
\bibitem{braemaud} 
  P.~Br\^{a}emaud, {\em Markov chains: Gibbs fields, Monte Carlo simulation, and queues},
  Springer, New York, 1999.
\bibitem{Metropolis:1953am}
  N.~Metropolis, A.~W.~Rosenbluth, M.~N.~Rosenbluth, A.~H.~Teller and E.~Teller,
  J.\ Chem.\ Phys.\  {\bf 21}, 1087 (1953).
\bibitem{Honda:2004yz}
  M.~Honda, T.~Kajita, K.~Kasahara and S.~Midorikawa,
  Phys.\ Rev.\ D {\bf 70}, 043008 (2004)
  [arXiv:astro-ph/0404457].
\bibitem{Gonzalez-Garcia:2004wg}
  M.~C.~Gonzalez-Garcia and M.~Maltoni,
  Phys.\ Rev.\ D {\bf 70}, 033010 (2004)
  [arXiv:hep-ph/0404085].
\bibitem{Maltoni:2001mt}
  M.~Maltoni, T.~Schwetz and J.~W.~F.~Valle,
  Phys.\ Lett.\ B {\bf 518}, 252 (2001)
  [arXiv:hep-ph/0107150].
\bibitem{MBrunplan}
  A.~A.~Aguilar-Arevalo {\it et al.} [MiniBooNE Collaboration],
  {\it The MiniBooNE Run Plan},
  {\tt http://www-boone.fnal.gov/publicpages/runplan.ps.gz}.
\bibitem{Casper:2002sd}
  D.~Casper,
  Nucl.\ Phys.\ Proc.\ Suppl.\  {\bf 112}, 161 (2002)
  [arXiv:hep-ph/0208030].
\bibitem{Maltoni:2002xd}
  M.~Maltoni, T.~Schwetz, M.~A.~Tortola and J.~W.~F.~Valle,
  Nucl.\ Phys.\ B {\bf 643}, 321 (2002)
  [arXiv:hep-ph/0207157].
\bibitem{Strumia:2002fw}
  A.~Strumia,
  Phys.\ Lett.\ B {\bf 539}, 91 (2002)
  [arXiv:hep-ph/0201134].
\bibitem{Grimus:2001mn}
  W.~Grimus and T.~Schwetz,
  Eur.\ Phys.\ J.\ C {\bf 20}, 1 (2001)
  [arXiv:hep-ph/0102252].
\bibitem{Hannestad:2005ey}
  S.~Hannestad,
  Prog.\ Part.\ Nucl.\ Phys.\ {\bf 57}, 309 (2006)
  [arXiv:astro-ph/0511595].
\bibitem{Abazajian:2004aj}
  K.~Abazajian, N.~F.~Bell, G.~M.~Fuller and Y.~Y.~Y.~Wong,
  Phys.\ Rev.\ D {\bf 72}, 063004 (2005)
  [arXiv:astro-ph/0410175].
\bibitem{Gelmini:2004ah}
  G.~Gelmini, S.~Palomares-Ruiz and S.~Pascoli,
  Phys.\ Rev.\ Lett.\ {\bf 93}, 081302 (2004)
  [arXiv:astro-ph/0403323].
\end{thebibliography}
\end{document}